\documentclass[fleqn,twoside]{article}
\usepackage{espcrc2}
\usepackage{amsmath,amssymb}

\usepackage{graphicx}
\usepackage[figuresright]{rotating}

\title{Selected challenges in low-energy QCD and hadron physics}

\author{Wolfram Weise\address[MCSD]{Physik-Department, Technische Universit\"at M\"unchen,        D-85747 Garching, Germany}%
\thanks{Work supported in part by BMBF, GSI and by the DFG Cluster of Excellence ``Origin and Structure of the Universe".}
       }
       
\begin{document}

\begin{abstract}
This presentation briefly addresses three basic issues of low-energy QCD: first, whether the Nambu-Goldstone scenario of spontaneous chiral symmetry breaking is well established; secondly, whether there is a dynamical entanglement of the chiral and deconfinement crossover transitions in QCD; and thirdly, what is the status of knowledge about the phase diagram of QCD at low temperature and non-zero baryon density.  These three topics were injected as key words into a panel discussion at the
Schladming school on Challenges in QCD. The following exposition reflects the style and character of the discussions, with no claim of completeness.  
\vspace{1pc}
\end{abstract}

\maketitle

\section{Topic I: \\Low-energy QCD with light quarks}

Is the Nambu-Goldstone spontaneous chiral symmetry breaking scheme confirmed in the low-energy
limit of QCD? 
The answer is positive: in the sector of light mesons, the effective field theory based on this scenario is in fact established as a quantitatively accurate science. 

One of the cornerstones in this development is the Gell-Mann, Oakes, Renner (GOR) relation,
\begin{equation}
m_\pi^2 = - {m_u + m_d\over 2f^2}\,\langle\bar{u}u + \bar{d}d\rangle + {\cal O}(m_q^2)~,
\label{gor} 
\end{equation}
where $m_{u,d}$ are the light quark masses, $\langle\bar{q}q\rangle$  with $q = u,d$ is the chiral (quark) condensate  and $f$ is the pion decay constant in the chiral limit. The chiral condensate, or alternatively, the pion decay constant act as order parameters for spontaneously broken chiral symmetry. The non-zero quark masses $m_{u,d}$ break chiral symmetry explicitly and move the pion from its massless Goldstone boson limit to its actual physical mass. Note that the quark masses and the chiral condensate are both renormalization scale dependent quantities whereas their product is scale invariant.

The question is now whether the proportionality $m_\pi^2 \propto m_q$, characteristic of Nambu-Goldstone symmetry breaking, is actually confirmed. Alternatively - as pointed out by J. Stern et al. long ago - the next-to-leading ${\cal O}(m_q^2)$ term  could in principle be sufficiently large so as to produce a linear relationship $m_\pi \propto m_q$ reminiscent of constituent quark models.

\begin{figure}[htb]
\includegraphics[width=7.5cm]{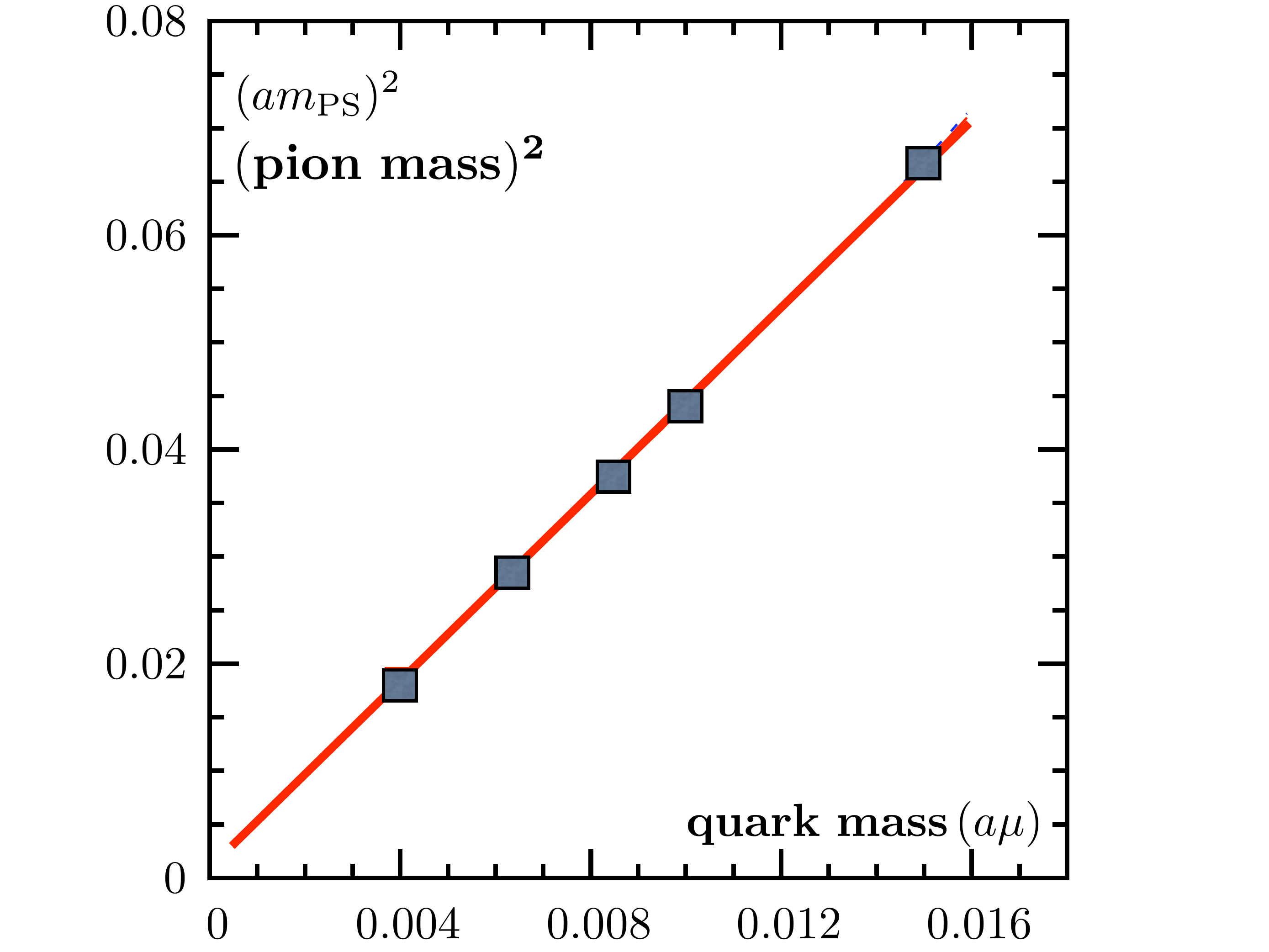}
\caption{Lattice QCD test \cite{ETM2007} of the GOR relation (\ref{gor}). Shown is the squared pion mass versus the quark mass, both in units of the lattice spacing $a$. The lowest lattice data point is at about twice the physical pion mass. }
\label{fig:1}
\end{figure}

Recent lattice QCD computations with light dynamical quarks together with precision data on low-energy $\pi\pi$ scattering now give consistent answers to this question \cite{ETM2007,EL2008}. Fig.\ref{fig:1} shows a lattice QCD test of the GOR relation. It demonstrates indeed that the leading piece in Eq.(\ref{gor}) holds up to about $m_\pi \simeq 500$ MeV.

%

\begin{figure}[htb]
\begin{minipage}[t]{7cm}
\includegraphics[width=7cm]{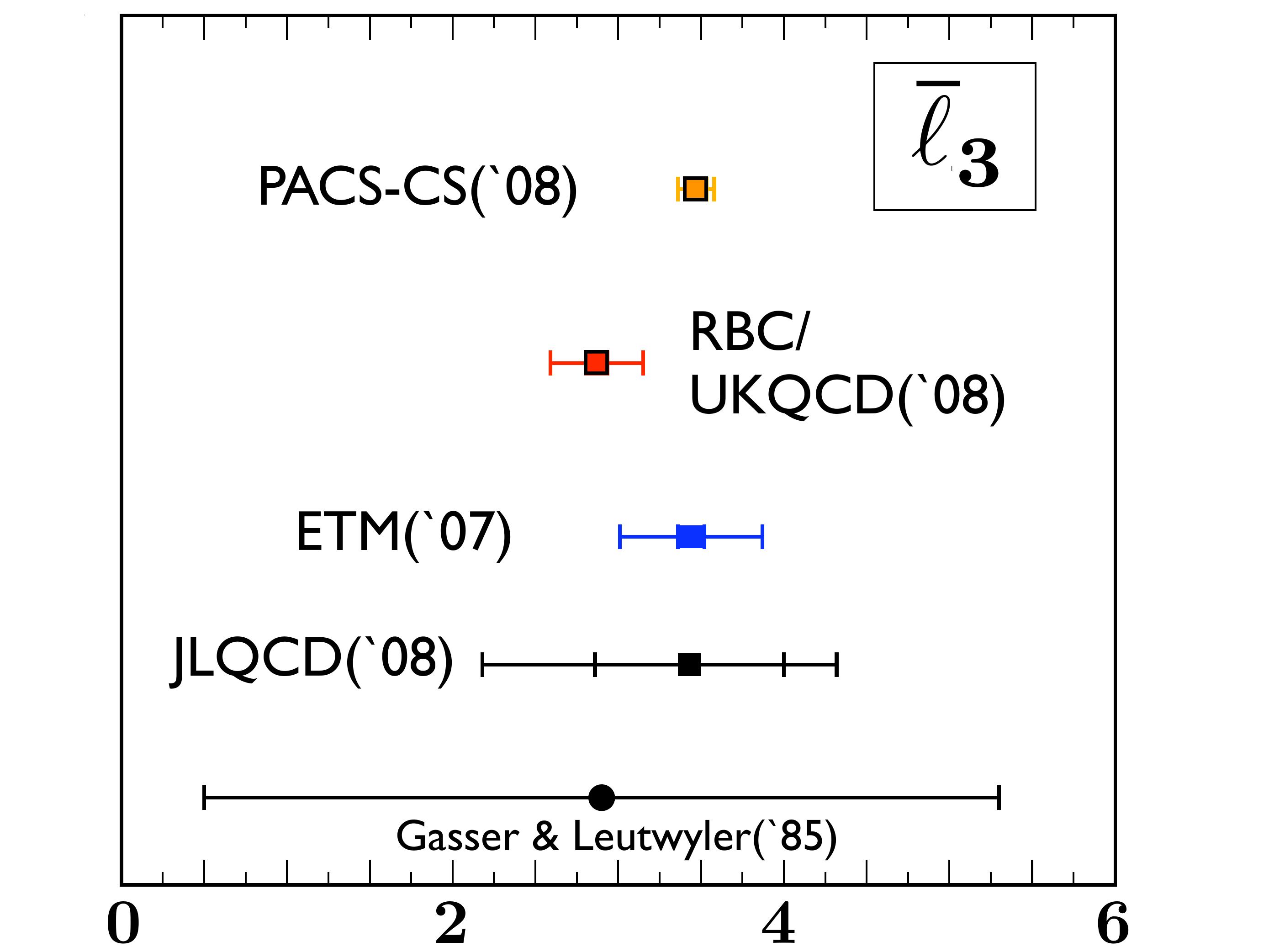}
\label{fig:2a}
\end{minipage}
\begin{minipage}[t]{7cm}
\includegraphics[width=7cm]{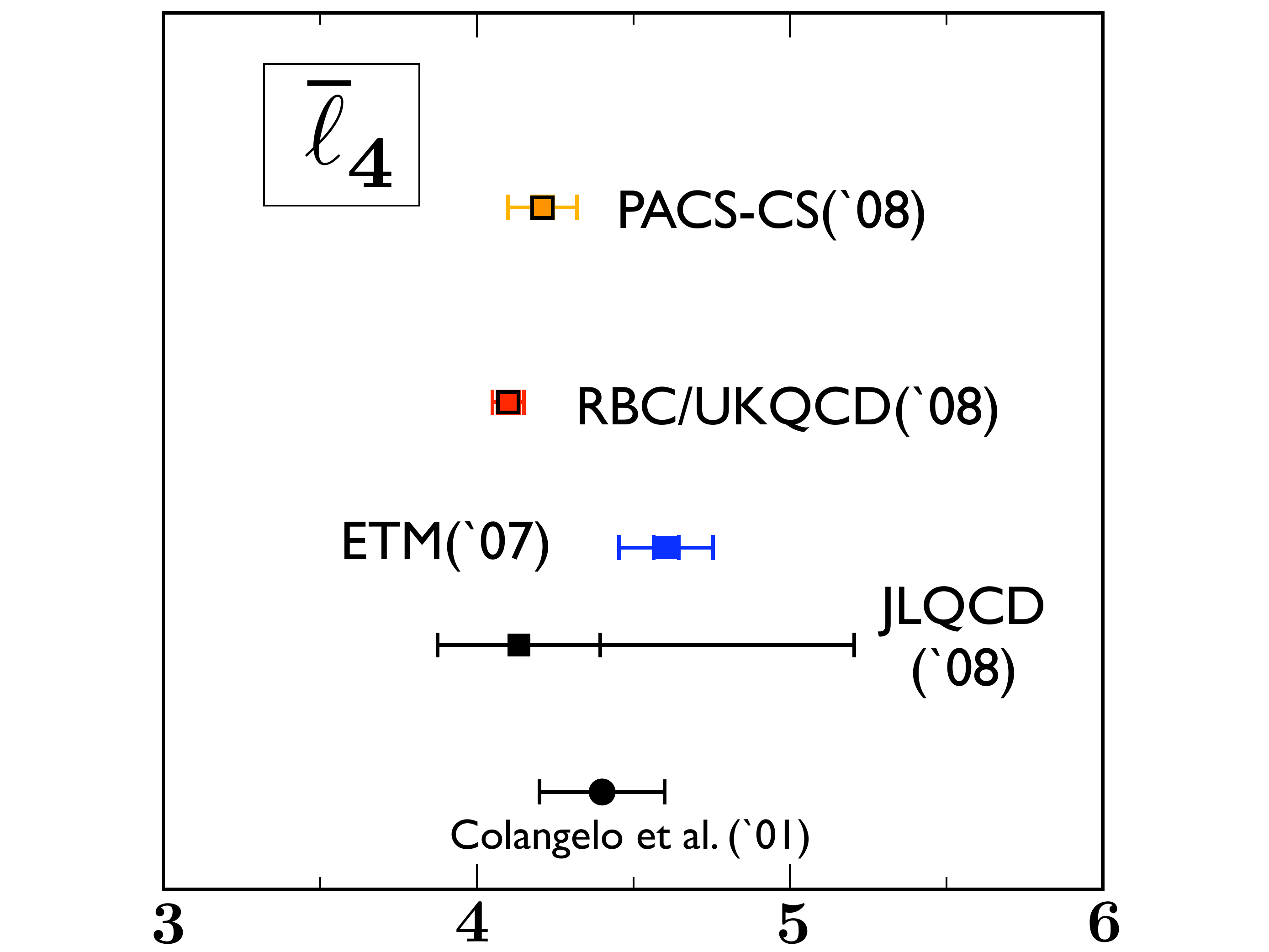}
\caption{Lattice QCD determinations of the low-energy constants $\bar{\ell}_3 = \ln(\Lambda_3^2/m_\pi^2)$ and   $\bar{\ell}_4=\ln(\Lambda_4^2/m_\pi^2)$. See Ref.\cite{EL2008} for further discussions.} 
\label{fig:2}
\end{minipage}
\end{figure}

Chiral perturbation theory gives the next-to-leading order corrections to Eq.(\ref{gor}) as follows:
\begin{eqnarray} 
m_\pi^2 &=& m^2 \left[1+{m^2\over 32\pi^2\,f^2}\ln{m^2\over\Lambda_3^2} +{\cal O}(m^4)\right]~, \\
f_\pi &=& f \left[1-{m^2\over 16\pi^2\,f^2}\ln{m^2\over\Lambda_4^2} +{\cal O}(m^4)\right]~,
\end{eqnarray}
where $m^2 = -(m_q/f^2)\langle\bar{u}u + \bar{d}d\rangle$ stands for the leading order. The NLO corrections involve two (renormalized) low-energy constants, $\bar{\ell}_3 = \ln(\Lambda_3^2/m_\pi^2)$ and   $\bar{\ell}_4=\ln(\Lambda_4^2/m_\pi^2)$. Lattice QCD determinations of these low-energy constants have now reached a remarkable level of precision, as evident from Fig.\ref{fig:2}. These results confirm that the NLO corrections to Eq.(\ref{gor}) are small as anticipated.  
The Nambu-Goldstone spontaneous symmetry breaking is thus indeed verified, with the non-perturbative QCD vacuum hosting a strong quark condensate.

\begin{figure}[htb]
\includegraphics[width=8cm]{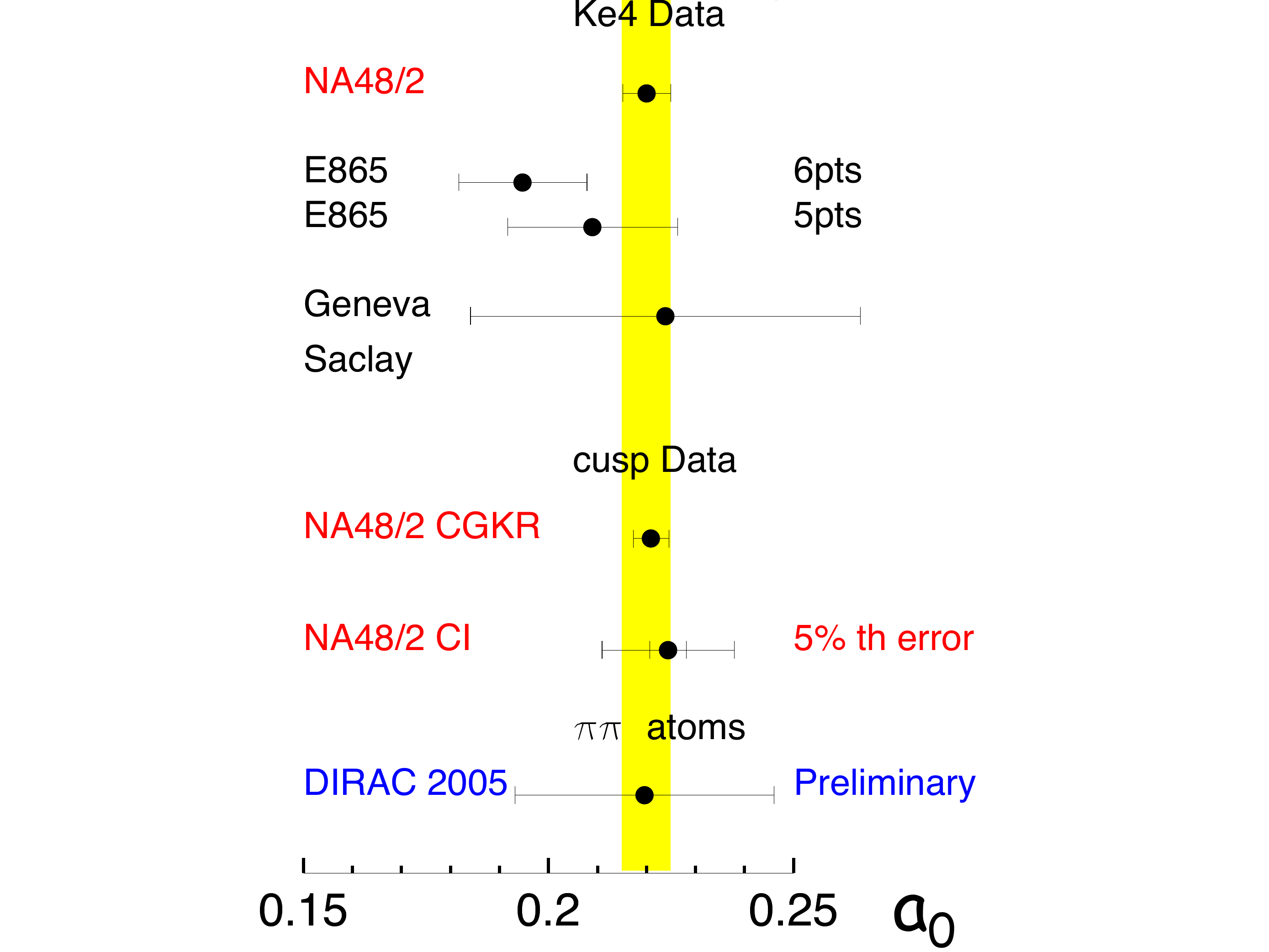}
\caption{Isoscalar $\pi\pi$ scattering length $a_0$ (in units of the inverse pion mass) deduced from Ke4 and $K\rightarrow 3\pi$ cusp data in comparison with the value calculated in chiral theory $(a_0 = 0.220\pm 0.005)$ \cite{Col2001}. Most recent high-precision data are provided by the CERN NA48/2 experiment. Also shown is a preliminary result from the DIRAC pionium measurement. Figure taken from Ref.\cite{BD2008}.} 
\label{fig:3}
\end{figure}

%
The s-wave $\pi\pi$ scattering lengths depend sensitively on the low-energy constants just mentioned. 
High precison data are available from the detailed analysis of the final state interaction in the Ke4 decay, $K^{\pm}\rightarrow \pi^+\pi^- e^{\pm}\nu(\bar{\nu})$, and from the cusp effect observed in near-threshold decays of kaons into three pions.  Results for the deduced isopin-zero $\pi\pi$ scattering length are summarized in Fig.\ref{fig:3} and shown to be in perfect agreement with theory, confirming the standard chiral symmetry breaking scenario. 

\begin{figure}[htb]
\begin{minipage}[t]{7cm}
\includegraphics[width=7cm]{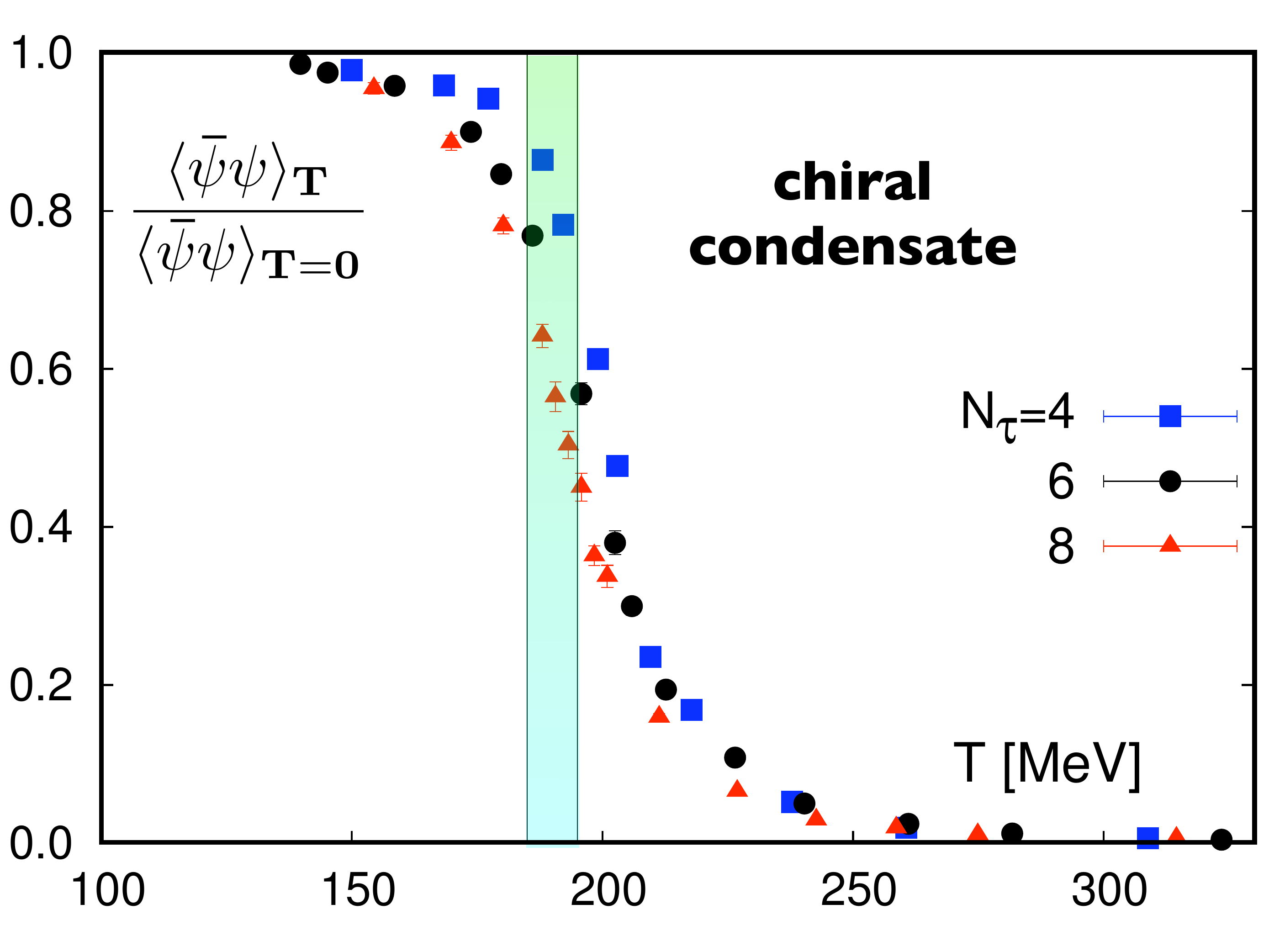}
\end{minipage}
\begin{minipage}[t]{7cm}
\includegraphics[width=7cm]{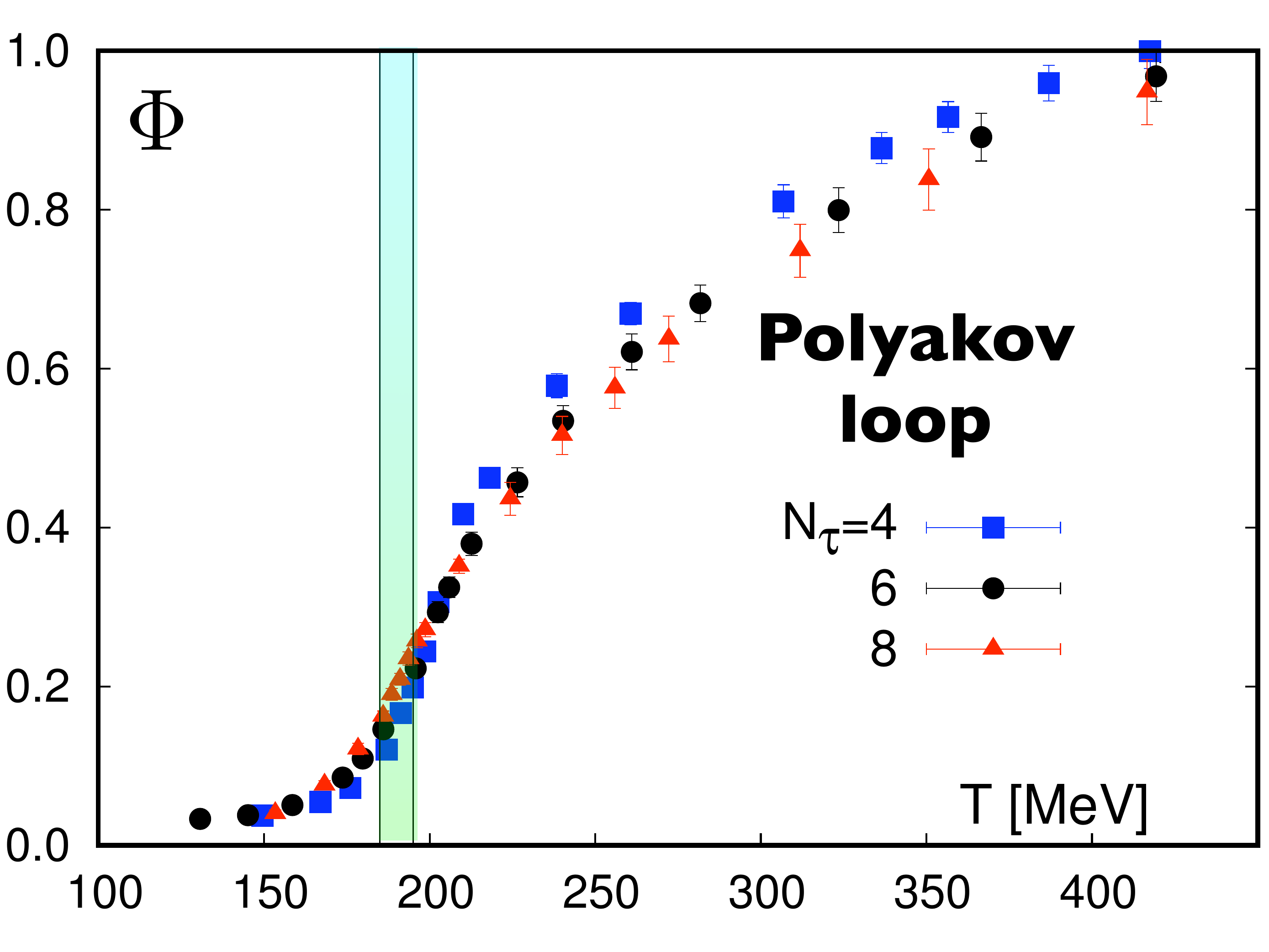}
\caption{Lattice QCD results ($N_f = 2+1$) of chiral condensate (upper part) und Polyakov loop (lower part) as functions of temperature \cite{Ch2008}. Different data sets correspond to different number $N_\tau$ of lattice point along the Euclidean time axis.} 
\label{fig:4}
\end{minipage}
\end{figure}
%

\section{Topic II:\\Deconfinement and chiral transitions}

The order parameter of spontaneously broken chiral symmetry is the quark condensate. The  disappearence of this condensate, by its melting above a transition temperature $T_c$, signals the restoration of chiral symmetry in Wigner-Weyl realization. 

The transition from confinement to deconfinement in QCD is likewise controlled by an order parameter, the Polyakov loop. A non-vanishing Polyakov loop $\Phi$ reflects the spontaneously broken $Z(3)$ symmetry characteristic of the deconfinement phase. The Polyakov loop vanishes in the low-temperature, confinement sector of QCD.

There is no principal reason why the deconfinement and chiral transitions should occur at the same temperature. Nonetheless, recent results of lattice QCD thermodynamics \cite{Ch2008} with 2+1 flavors (at zero baryon chemical potential) indicate just that (see Fig.\ref{fig:4}), with a common transition temperature $T_c \simeq 190$ MeV. A deeper understanding of this observation is of fundamental importance, also in view of the fact
that earlier lattice simulations \cite{Aoki2006} still found a displacement between chiral and deconfinement temperatures (although a more detailed assessment of systematic uncertainties might resolve this apparent contradiction).

One notes that the chiral and deconfinement transitions as shown in Fig.\ref{fig:4} are not phase transitions but smooth crossovers, so there is no critical temperature in the strict sense. It is nevertheless possible to define a transition temperature band around the maximum slope of either the condensate $\langle\bar{q}q\rangle_T$ or the Polyakov loop $\Phi(T)$. 

Two limiting cases are of interest in this context. In pure gauge QCD, corresponding to infinitely heavy quarks, the deconfinement transition is established - at least in lattice QCD - as a first order phase transition with a critical temperature of about $270$ MeV. In the limit of massless $u$ and $d$ quarks, on the other hand, the isolated chiral transition appears as a second order phase transition at a significantly lower critical temperature. This statement is based on calculations using Nambu - Jona-Lasinio (NJL) type models which incorporate the correct spontaneous chiral symmetry breaking mechanism but ignore confinement. The step from first or second order phase transitions to crossovers is understood as a consequence of explicit symmetry breaking. The $Z(3)$ symmetry is explicitly broken by the mere presence of quarks with non-infinite masses. Chiral symmetry is explicitly broken by non-zero quark masses. But the challenging question remains how the chiral and deconfinement transitions get dynamically entangled in just such a way that they finally occur at a joint transition temperature interval.

Insights concerning this issue can be gained from a model based on a minimal synthesis of the NJL-type spontaneous chiral symmetry breaking mechanism and confinement implemented through Polyakov loop dynamics. This PNJL model \cite{Fu2003,RTW2006} is specified by the following action:
\begin{eqnarray}
{\cal S} = \int_0^{\beta=1/T}d\tau \int_V d^3x\left[\psi^\dagger\partial_\tau\psi-{\cal H}(\psi,\psi^\dagger,\phi)\right] \nonumber\\ - {V\over T}\,{\cal U}(\Phi,T)~.\nonumber
\end{eqnarray}
It introduces the Polyakov loop, $\Phi = N_c^{-1}\,Tr\exp(i\phi/T),$ with a homogeneous temporal gauge field, $\phi = \phi_3\lambda_3 + \phi_8\lambda_8\in SU(3)$. Its dynamics is controlled by a $Z(3)$ symmetric effective potential ${\cal U}$, designed such that it reproduces the equation of state of pure gauge lattice QCD with its first order phase transition at a critical temperature of 270 MeV. The field $\phi$ acts as a potential on the quarks represented by the flavor doublet (for $N_f = 2$) or triplet (for $N_f = 3$) Fermion field $\psi$. The coupling of $\phi$ to the quark density $\psi^\dagger\psi$ is dictated by color gauge invariance. The Hamiltonian density in the quark sector is ${\cal H} = -i\psi^\dagger(\vec{\alpha}\cdot\vec{\nabla} +\gamma_4\,\hat{m} - \phi)\psi + {\cal V}(\psi,\psi^\dagger)$, with the quark mass matrix $\hat{m}$ and a chiral $SU(N_f)_L\times SU(N_f)_R$ symmetric interaction ${\cal V}$. 

Earlier two-flavor versions of the PNJL model \cite{Fu2003,RTW2006,RRW2007} have still used a local four-point interaction of the classic NJL type, requiring a momentum space cutoff to regularize loops. A more recent version \cite{HRCW2009} using a non-local interaction does not require any artificial cutoff. It generates instead a momentum dependent dynamical quark mass, $M(p)$, along with the non-vanishing quark condensate. A further extension to three flavors includes a $U(1)_A$ breaking term implementing the axial anomaly of QCD. 

\begin{figure}[htb]
\includegraphics[width=7cm]{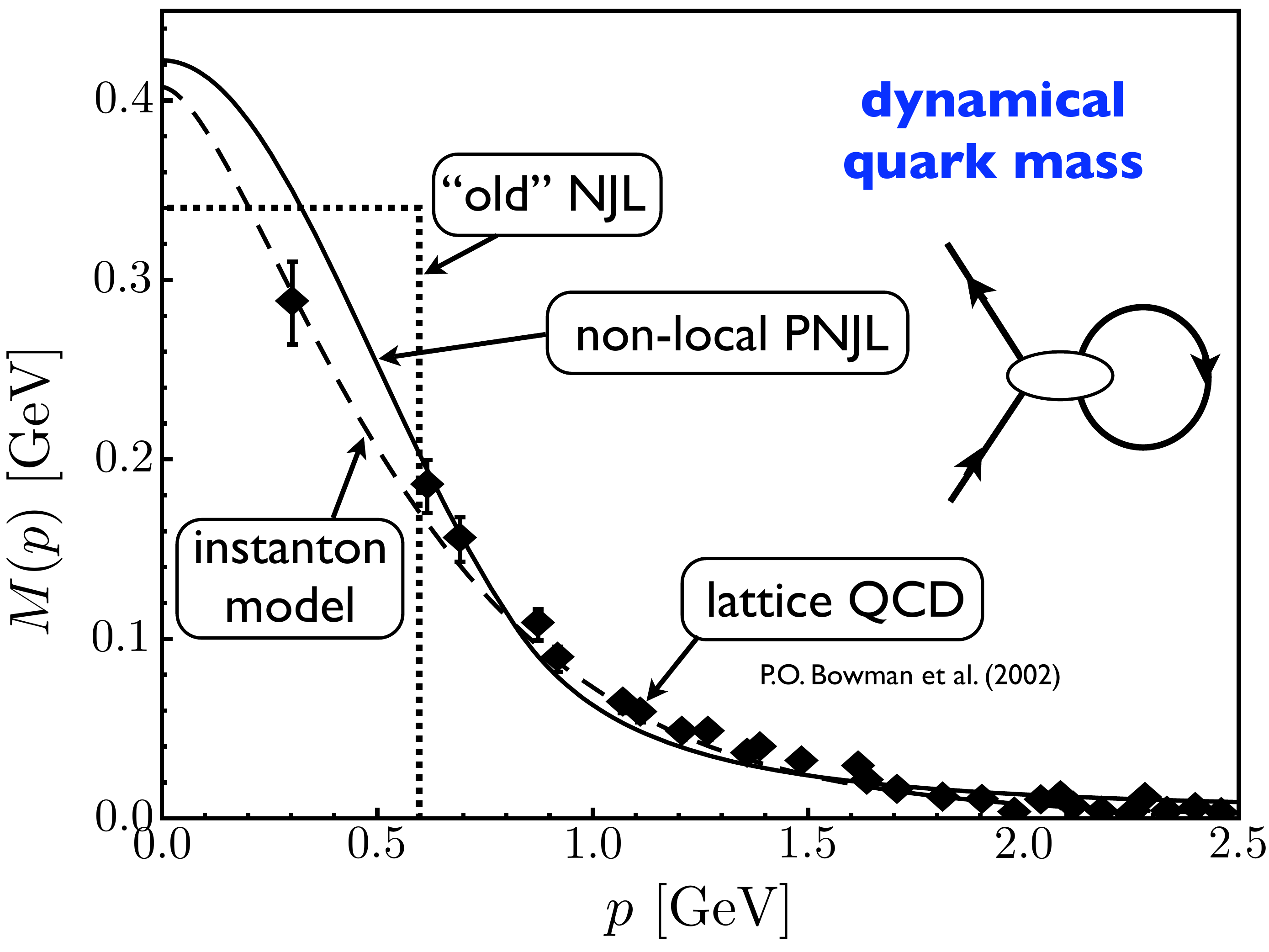}
\caption{Dynamical quark mass $M(p)$ resulting from the gap equation of the non-local PNJL model (solid line). The dashed line shows a similar mass profile produced with an interaction kernel from an instanton model. Lattice QCD data (extrapolated to physical $u$ and $d$ quark masses) are shown for comparison, as well as the step function mass profile (with cutoff) of the classic NJL model. (Figure adapted from Ref.\cite{HRCW2009}.)} 
\label{fig:5}
\end{figure}

The non-local approach leads to a self-consistent gap equation
\begin{eqnarray}
M(p) = m_0 + 4N_f N_c \int {d^4q\over (2\pi)^4} G(p-q){M(q)\over q^2 + M^2(q)} \nonumber
\end{eqnarray}
for the dynamically generated quark mass, $M(p)$, starting from a small (current) quark mass $m_0$. A typical result is shown in Fig.\ref{fig:5}. The running coupling strength of the non-local interaction ${\cal V}$ represented by the distribution $G(p)$ is chosen such that it matches QCD constraints on the quark mass function at high momentum and extrapolated lattice results at lower momentun, or the quark self-energy from Dyson-Schwinger calculations using Landau gauge. Using an instanton model is yet another option. The pseudoscalar meson spectrum together with the pion and kaon decay constants at zero temperature are well reproduced \cite{HRCW2009}. 

With the input fixed at zero temperature, the thermodynamics of the PNJL model can then be investigated with focus on the symmetry breaking pattern and on the intertwining of chiral dynamics with that of the Polyakov loop. The primary role of the Polyakov loop and its coupling to the quarks is to cut down the residues of the thermal quark and diquark propagators, those that are non-singlets in color, when the 
transition temperature $T_c$ is approached from above. Color singlets, on the other hand,  are left to survive below $T_c$.
\begin{figure}[htb]
\begin{minipage}[t]{7cm}
\includegraphics[width=7cm]{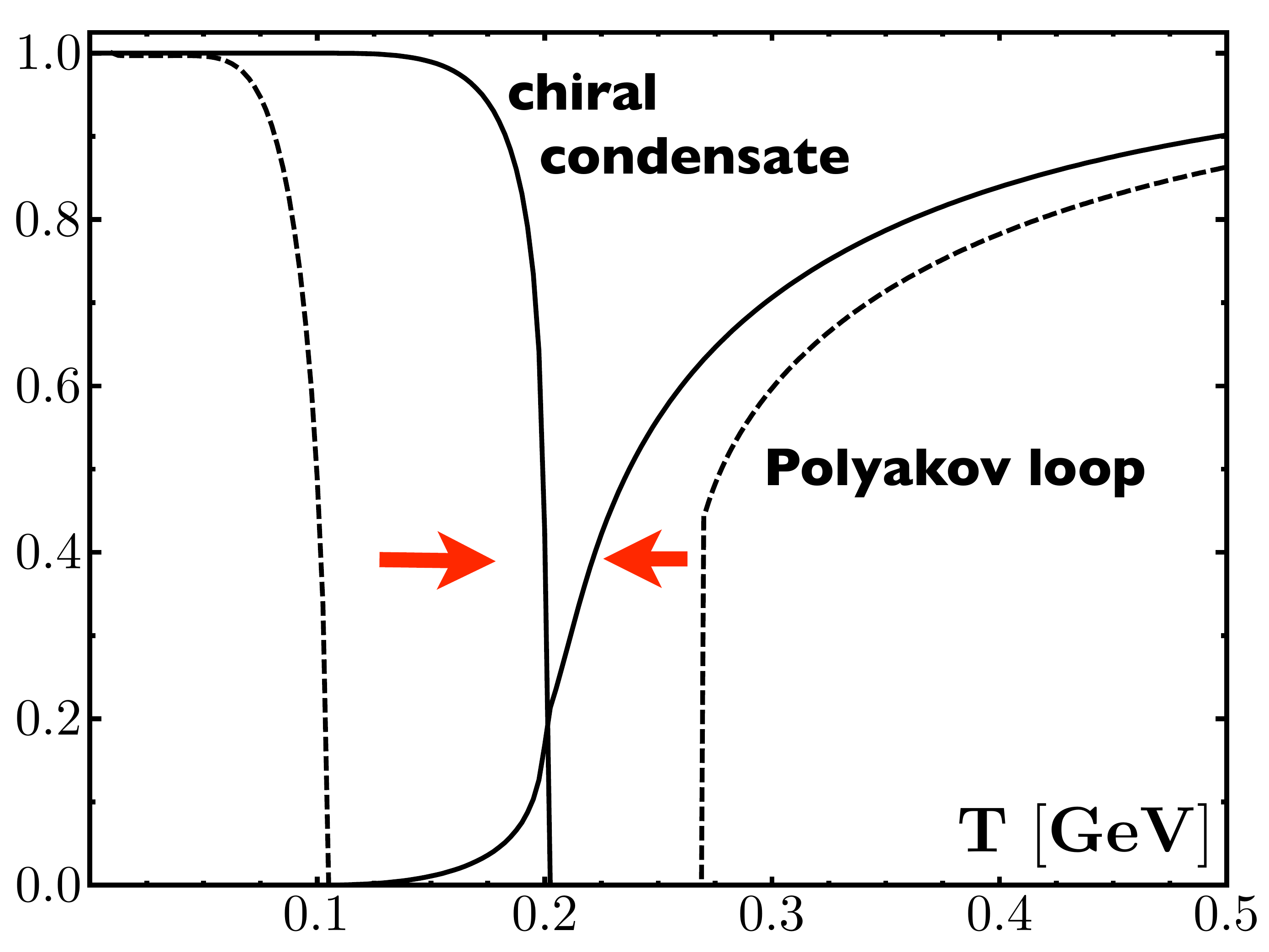}
\label{fig:6a}
\end{minipage}
\begin{minipage}[t]{7cm}
\includegraphics[width=7.3cm]{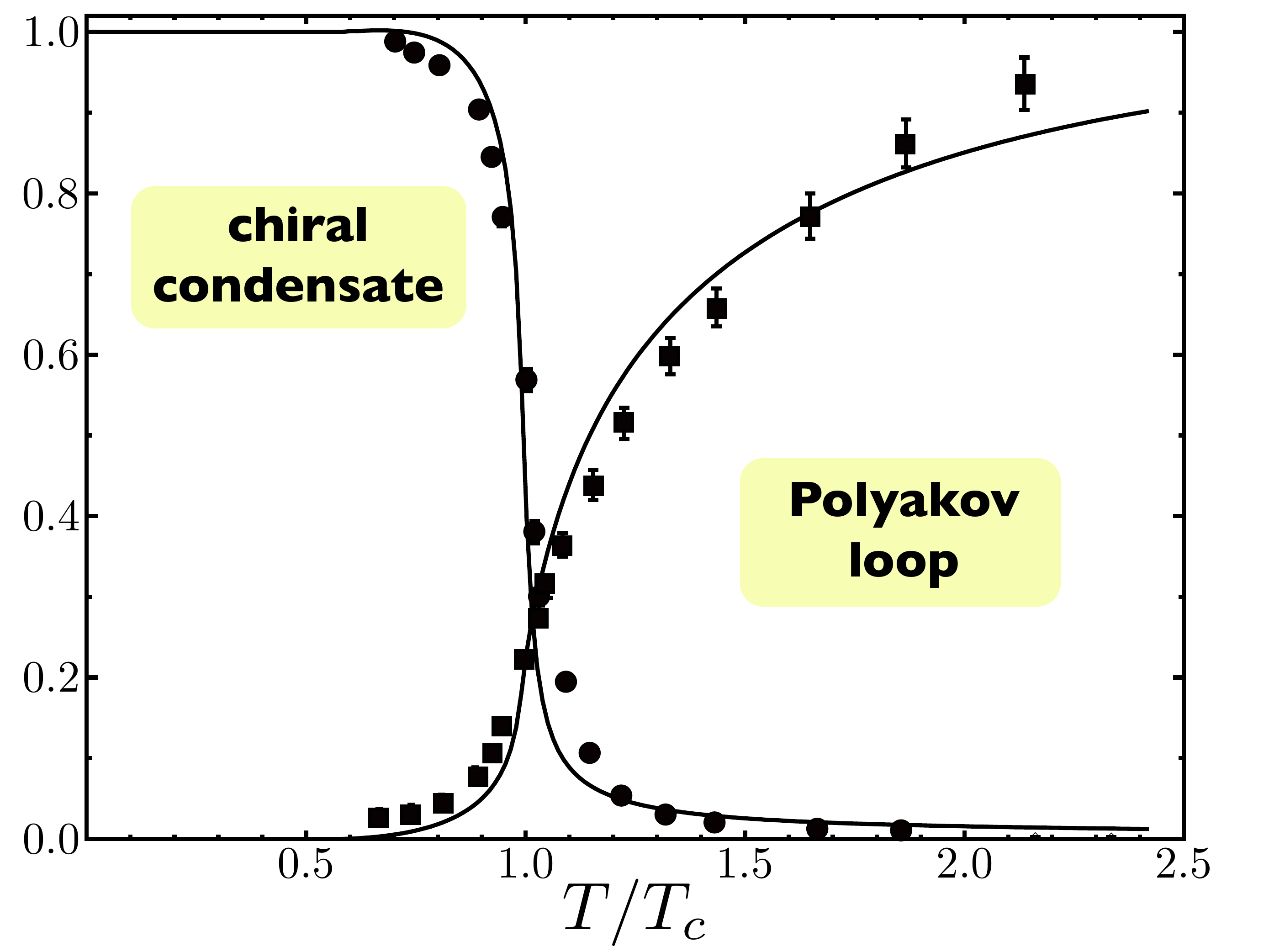}
\caption{PNJL model calculations of chiral and deconfinement transitions \cite{RRW2007,HRCW2009}. See text for  explanations.}
\label{fig:6}
\end{minipage}
\end{figure}
%

A remarkable dynamical entanglement of the chiral and deconfinement transitions is then observed, as demonstrated in Fig.\ref{fig:6} (upper) for the two-flavor case. In the absence of the Polyakov loop the quark condensate (left dashed line), taken in the chiral limit, shows the expected 2nd order chiral phase transition, but at a temperature way below and far separated from the 1st order deconfinement transition controlled by the pure-gauge Polyakov loop effective potential ${\cal U}$ (right dashed line). Once the coupling of the Polyakov loop field to the quark density is turned on, the two transitions move together and end up at a common transition temperature around 0.2 GeV. The deconfinement transition has turned into a crossover (with $Z(3)$ symmetry explicitly broken by the coupling to the quarks), while the chiral phase transition remains 2nd order until non-zero $u$ and $d$ quark masses, $m_{u,d}  \simeq 4$ MeV,  induce a crossover transition as well.  

The lower part of Fig.\ref{fig:6} shows the two-flavor PNJL result \cite{HRCW2009} together with $N_f=2+1$ lattice data \cite{Ch2008}. A direct comparison is clearly not appropriate but the similarity of the crossover transition patterns is striking, given the simplicity of the model and the fact that these results are derived from a mean-field approach. A further important step is now to systematically investigate effects beyond mean field.

\section{Topic III:\\Scenarios at finite baryon density}

\begin{figure}[htb]
\includegraphics[width=8.5cm]{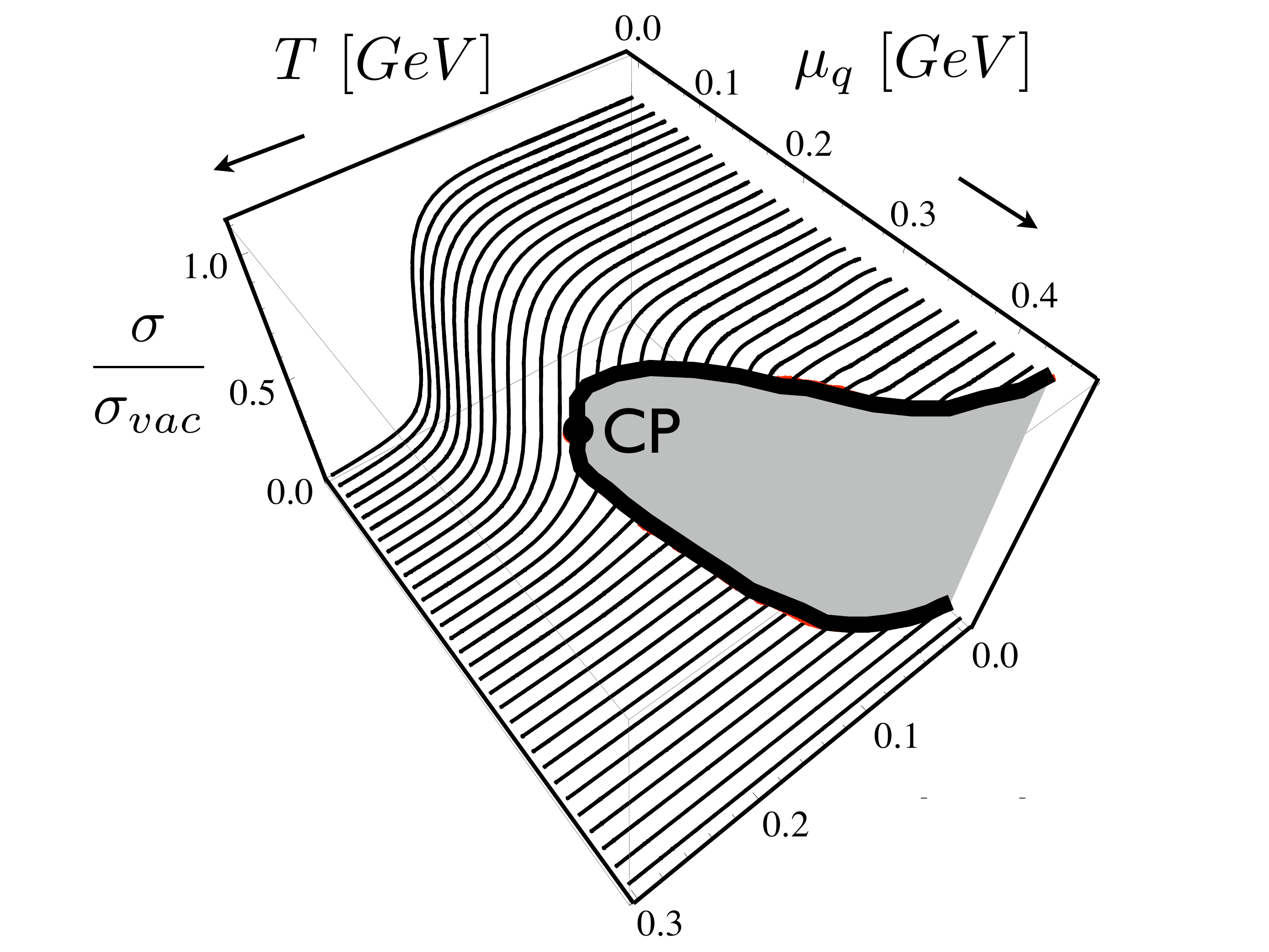}
\caption{PNJL model calculation \cite{HRCW2009} of the scalar field representating the chiral condensate as a function of temperature $T$ and quark chemical potential $\mu_q$, showing a first order phase transition with critical point CP. } 
\label{fig:7}
\end{figure}

Undoubtedly a prime challenge in the physics of strong interactions is the exploration of the QCD phase diagram at non-zero baryon density, extending from normal nuclear matter all the way up to the large quark chemical potentials $\mu$ at which color superconducting phases presumably occur. PNJL calculations at finite $\mu$ give a typical pattern of the chiral order parameter as presented in Fig.\ref{fig:7}, showing a first order transition with a critical point (CP). 

Several important questions are being raised in this context. The first one concerns the existence and location of the critical point. Extrapolations from lattice QCD, either by Taylor expansions around $\mu =0$
\cite{FK2006} or by analytic continuation from imaginary chemical potential \cite{FP2008}, have so far not reached a consistent conclusion. A second question relates to the sensitivity  of the first order transition line in the phase diagram with respect to the axial $U(1)_A$ anomaly in QCD. This issue has been addressed in Ref.\cite{YHB2007} and it was pointed out that, depending on details of the axial $U(1)_A$ breaking interaction, a second critical point might appear such that the low-temperature evolution to high density is again just a smooth crossover, or the first order transition might disappear altogether and give way to a smooth crossover all over the place.

\begin{figure}[htb]
\includegraphics[width=8cm]{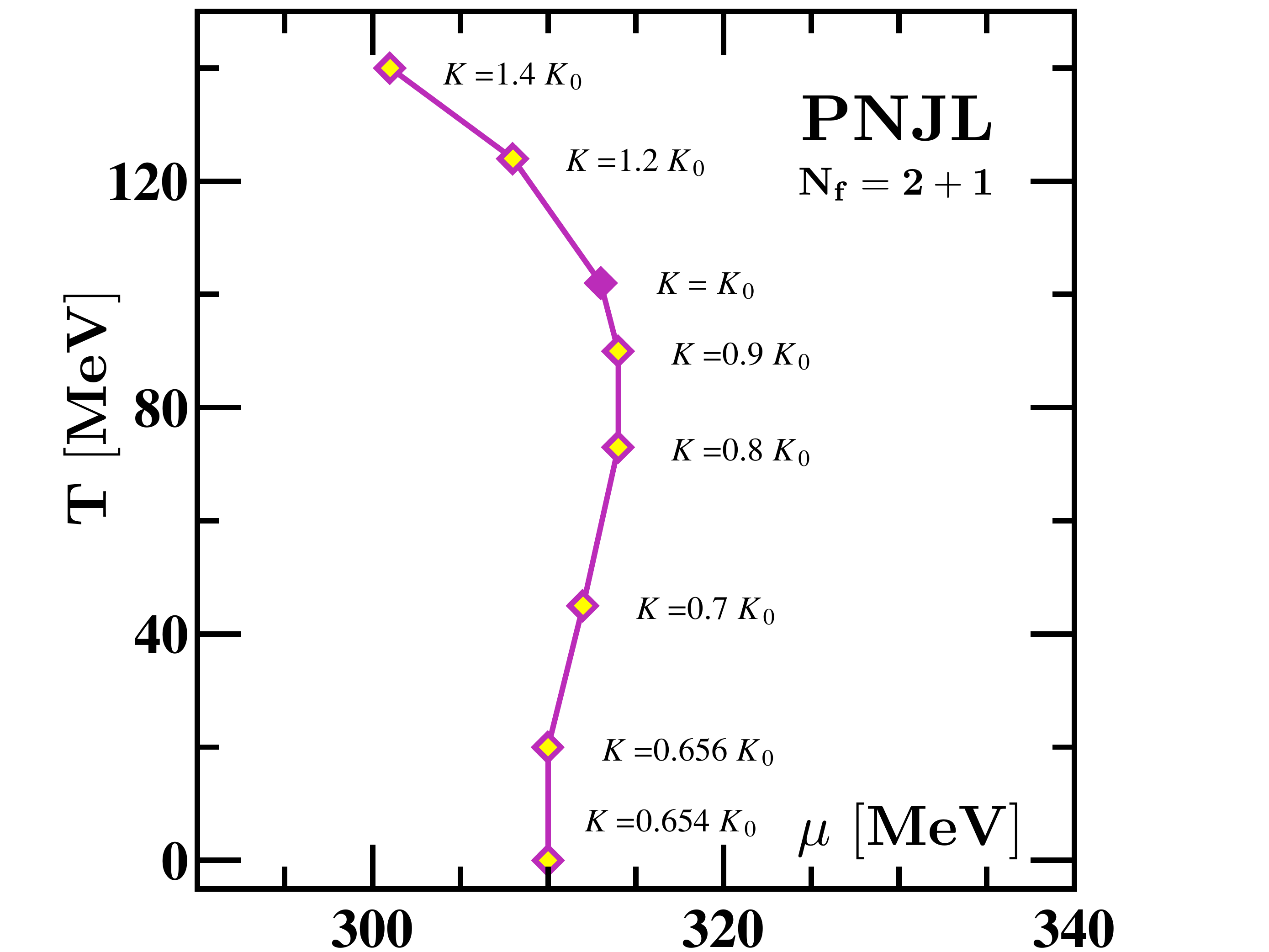}
\caption{Three-flavour PNJL calculation \cite{BHRW2009} showing the dependence of the critical point on the strength $K$ of the axial $U(1)$ breaking interaction; $K=K_0$ is the point at which the $\eta'$ mass in vacuum is reproduced.} 
\label{fig:8}
\end{figure}

An impression of the explicit dependence of the critical point on the axial anomaly can be obtained using the three-flavor PNJL model with inclusion of a $U(1)_A$ breaking Kobayashi-Maskawa-`t Hooft determinant interaction and varying the coupling strength $K$ of this interaction \cite{Fu2008,BHRW2009}. It turns out that the location of the critical point in the phase diagram varies indeed strongly with $K$ and may even disappear altogether below a certain value of $K$ (see Fig.\ref{fig:8}). A closely related question is how the mass of the $\eta\,'$ meson behaves in a dense baryonic environment. 

From the variety of existing model calculations (including those using the PNJL model) one might draw the presumably premature conclusion that critical phenomena occur already at densities not much higher than that of normal nuclear matter. However, all these models are not capable of working with the proper degrees of freedom around and below a baryon chemical potential of 1 GeV (corresponding to quark chemical potentials around 0.3 GeV). Approaching this density scale from below, it is obvious that constraints from what we know about the nuclear matter equation of state must be seriously considered.

\begin{figure}[htb]
\includegraphics[width=7.5cm]{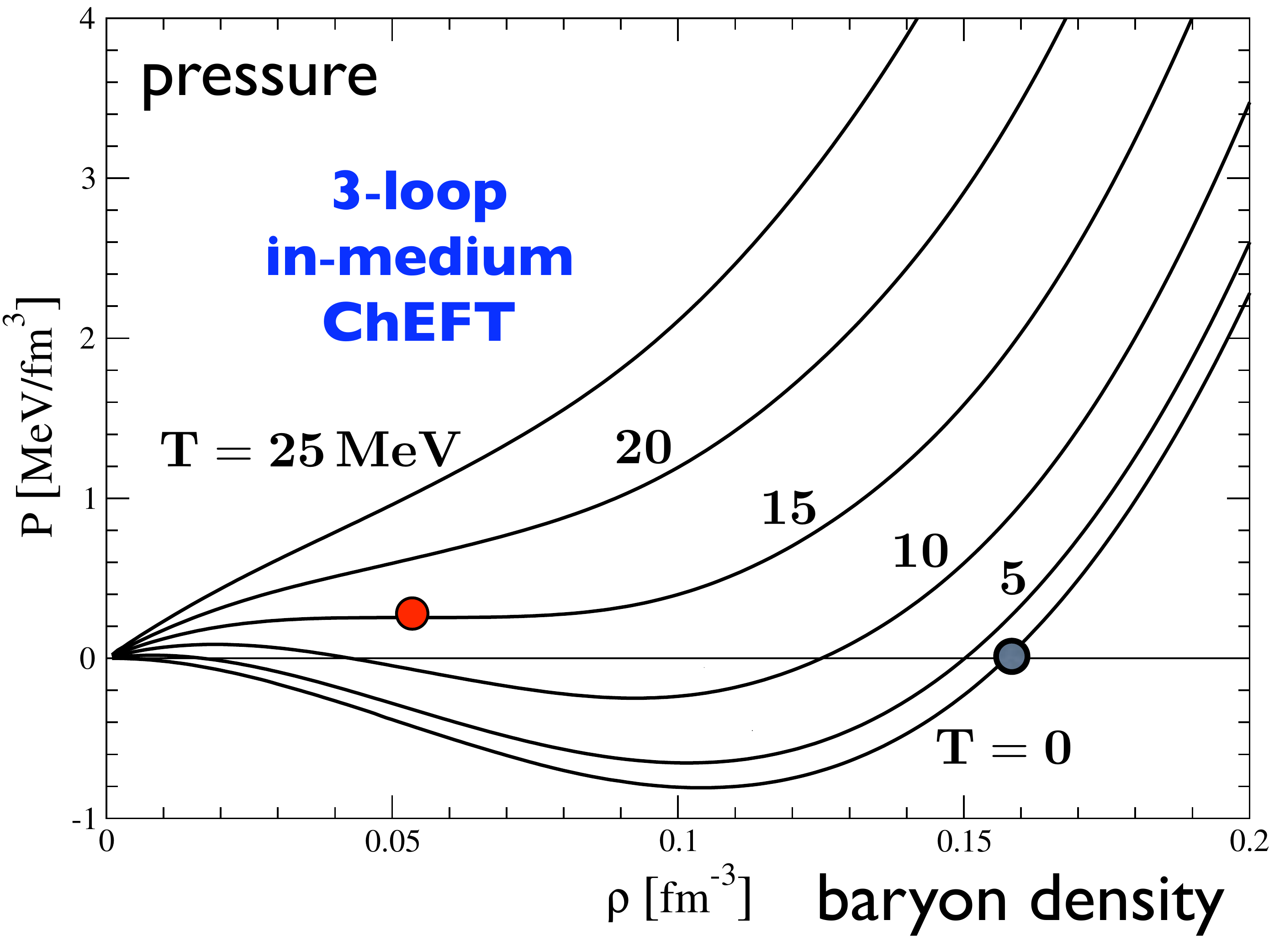}
\caption{Nuclear mater equation of state derived from in-medium chiral effective field theory at three-loop order \cite{FKW2005}.} 
\label{fig:9}
\end{figure}
Chiral effective field theory is not only the low-energy realization of QCD in the meson and single-baryon sectors. It is also a basis for dealing with the nuclear many-body problem in terms of in-medium chiral perturbation theory \cite{FKW2005}. In this approach, chiral one- und two-pion exchange processes in the nuclear medium are treated explicitly while unresolved
short-distance dynamics is encoded in contact terms. Present calculations are performed up to three-loop order in the energy density. Three-body interactions emerge and play a significant role in this framework. The pion mass $m_\pi$, the nuclear Fermi momentum $p_F$ and the
mass splitting between nucleon and $\Delta(1232)$ are all comparable scales and so the relevant,
active degrees of freedom are pions, nucleons and $\Delta$ isobars. Intermediate range two-pion exchange interactions produce van der Waals - like forces involving the large spin-isospin polarizablity
of the individual nucleons. The Pauli principle acts on intermediate nucleon states in two-pion exchange
processes. 

This scenario leads to a realistic nuclear matter equation of state as presented in Fig.\ref{fig:9}, with a liquid-gas first order phase transition and a critical temperature of about 15 MeV, close to the range of empirical values extracted for this quantity. We show this figure here primarily because it draws a picture of a well established part of the phase diagram of strongly interacting matter at finite density and low temperature. Given the truncation in the chiral expansion of the pressure in powers of the Fermi momentum $p_F$, these results can probably be trusted in the range $p_F \lesssim 0.3$ GeV, corresponding to baryon densities up to about twice the density of normal nuclear matter. 

Equiped with such a nuclear equation of state based on chiral dynamics in which the pion mass (or, equivalently, the quark mass according to the GOR relation (\ref{gor})) enters explicitly, one can now ask the following question: how does the chiral condensate extrapolate
to baryon densities exceeding those of normal nuclear matter? In-medium chiral effective field theory
gives a definite answer:
\begin{eqnarray}
{\langle\bar{q}q\rangle_\rho\over\langle\bar{q}q\rangle_0} = 1&-&{\rho\over f_\pi^2}{\sigma_N\over m_\pi^2}\left(1 - {3p_F^2\over 10M_N^2} + \dots\right) \nonumber\\
&+&{\rho\over f_\pi^2}{\partial\over \partial m_\pi^2}\left({E_{int}(p_F)\over A}\right)~. 
\end{eqnarray}
The first term with its leading linear dependence on density is the contribution from a free Fermi gas of nucleons, with the pion-nucleon sigma term $\sigma_N\simeq 0.05$ GeV and non-static corrections. The term in the second line involves the pion mass dependence of the interaction energy per nucleon, $E_{int}/A$. This term features prominently the two-pion exchange interaction in the nuclear medium  including Pauli principle corrections, and also three-nucleon forces based on two-pion exchange.
\begin{figure}[htb]
\includegraphics[width=7.5cm]{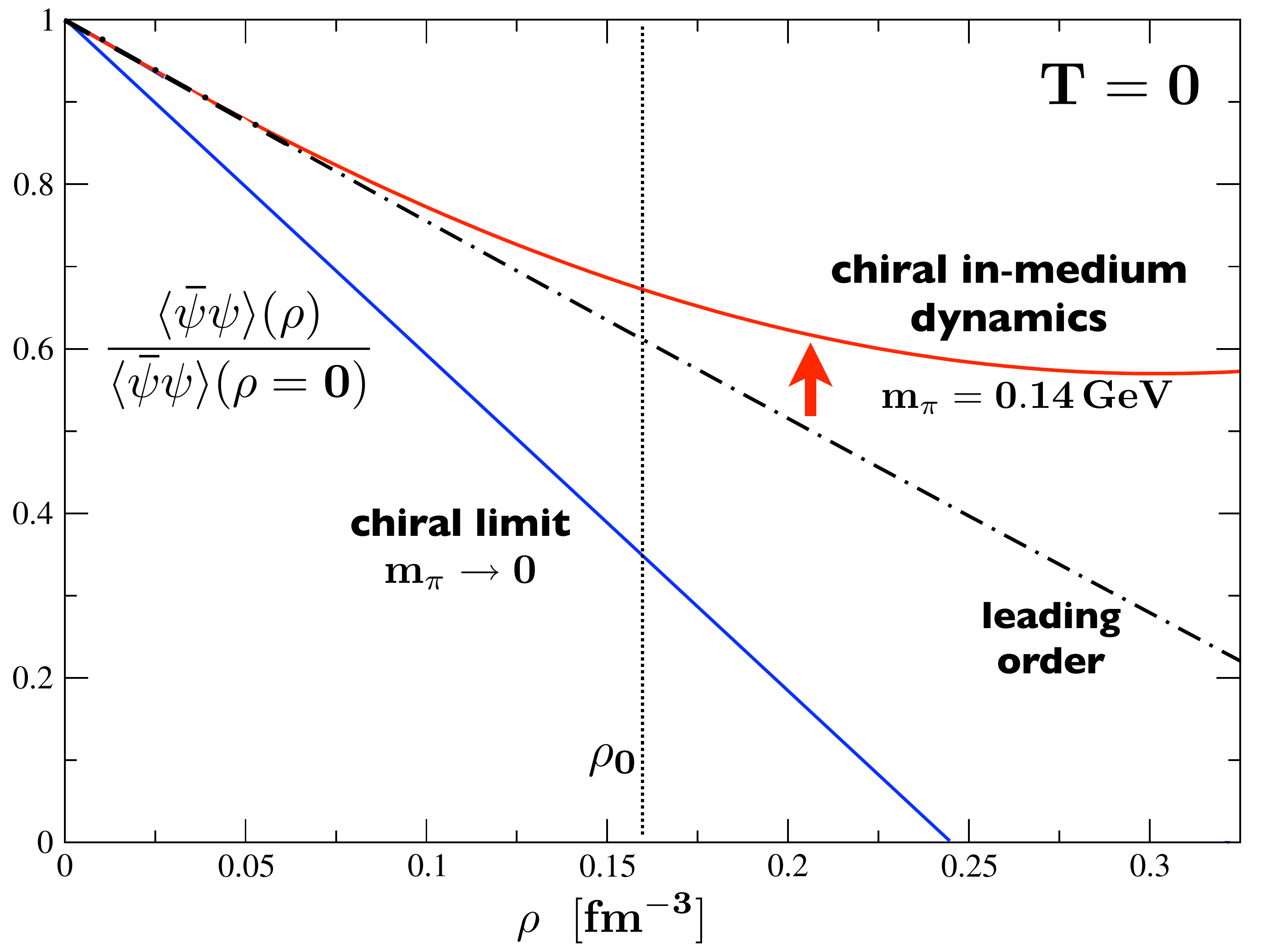}
\caption{Density dependence of the chiral condensate in symmetric nuclear matter \cite{KHW2008}. Dot-dashed curve: leading order term using $\sigma_N = 50$ MeV. Upper curve: full in-medium chiral dynamics result at three-loop order. Lower curve: chiral limit with vanishing pion mass.} 
\label{fig:10}
\end{figure}

The dashed-dotted curve in Fig.\ref{fig:10} shows the pronounced leading linear reduction in the magnitude of the chiral condensate with increasing density. This holds in the absence of correlations between the nucleons. Up to about the density of normal nuclear matter, this term dominates, whereas the interaction part tends to delay the tendency towards chiral restoration when the baryon density is further increased. This behaviour is sensitive to the actual value of the pion mass. In the chiral limit, $m_\pi\rightarrow 0$, with stronger attraction in the NN force at intermediate ranges, the trend is reversed and the rapidly dropping condensate would now lead to the restoration of chiral symmetry at relatively low density: nuclear physics would look completely different if the pion were an exactly massless Nambu-Goldstone boson. The influence of  explicit chiral symmetry breaking through the small but non-zero $u$ and $d$ quark masses on qualitative properties of nuclear matter is quite remarkable. 

These results demonstrate that known properties of a realistic nuclear equation of state must be considered as important constraints for extrapolations to higher densities, at least at low temperatures.
PNJL type models work with quarks as baryonic quasiparticles.  They do not account for those parts of the QCD phase diagram that prominently involve color singlet baryons as relevant degrees of freedom. 
The recent very interesting discussion  \cite{McL2009} of a quarkyonic sector in the phase diagram at moderate quark chemical potentials should not miss these constraints.\\

{\bf Acknowledgements}\\

Many thanks to Leonid Glozman, Christof Gattringer and their colleagues in Graz for arranging the 2009 Schladming School as a most inspiring event. Stimulating discussions with Kenji Fukushima are gratefully acknowledged. Special thanks  go also to my present and former collaborators Nino Bratovic, Marco Cristoforetti, Thomas Hell, Norbert Kaiser, Bertram Klein, Claudia Ratti and Simon R\"o{\ss}ner whose works have contributed substantially to this report.\\

\end{document}